The O(oxygen)-M(metal)-O(oxygen) molecule is a basic unit of high-temperature superconducting cuprates and colossal magnetoresistance exhibiting manganates. This molecule can be regarded either as an element of a linear chain or as an ingredient of the corresponding cuprate or manganate lattice. The symmetry of the unit being different in the two approaches, group theory imposes different limitations on conceivable vibrational modes and atomic orbitals that control its transport and optical properties. We now calculate the electron hopping energies along Cu(P)-O(A) bonds, sites for nonlocal electron-vibrational mode coupling. We find the electric transport along the O(A)-Cu(P)-O(A) molecule dominated by scattering from bond polarons which is reflected in the two-branch character of the temperature dependence of its electric resistance.

Literature reference. There are two prerequisites of superconductivity: pairing and long-range phase coherence via interlayer (axial) coupling. Path-integral analyses suggest a coherent axial transport in normal state dominated by the scattering of fermionic excitations by double-well potentials associated with apex oxygens.[1] However, magnetoresistance data indicate that the axial transport in $La_{2-x} Sr_x CuO_4$ is largely incoherent, carriers undergoing significant in-plane scattering before hopping to another conducting plane.[2] The axial leak is still an open issue.

The electric resistivity $\rho$ in anisotropic high-temperature superconductors (HTSC) is different in ab-plane and along the c-axis. Both $\rho_{ab}$ and $\rho_c$ are measured as functions of temperature T and doping *x* in crystals with carrier concentration and quantum state controlled by chemical substitution and external magnetic field.[3,4] The temperature dependencies of $\rho_{ab}$ and $\rho_c$ exhibit common trends as T is increased: a semiconductor-like decrease ($d\rho/dT<0$) at lower T followed by a metal-like increase ($d\rho/dT>0$) at higher T.

Rationale of present work. We derive a temperature dependence for the resistivity assuming strong electron-phonon coupling leading to polaron formation. A drift velocity is defined as the difference forth and back of the phonon-coupled electron transfer rate along the axial Cu(P)-O(A) bond. The resulting temperature dependence is in concert with domains of experimental data on high-$T_c$ oxides.

This conclusion is independent of the form of the polaron potential. But in order to make model closer to experiment,[5] we consider double-well polaron potentials splitting the apex oxygen site. We also make a guess on the symmetry of the coupled vibration as $A_{2u}$ or $A_{2g}$ and $E_u$ or $E_g$, axial and in-plane, depending on the symmetry of electronic states involved.

Electric transport along Cu(P)-O(A) bond. For deriving a current, we define a *drift velocity* along the Cu(P)-O(A) bond, namely:

$$J_c(T) = Ne\, L\, [\, R_+(T,\mathbf{F}) - R_-(T,\mathbf{F})\, ]$$

(N - carrier concentration, e - electron charge, L - bond length). $R_\pm(T,\mathbf{F})$ are the temperature-(T) and field-($\mathbf{F}$) dependent rates of electron transfer along the bond, forth (+) and back (-). The phonon-coupled rate in an external electric field $\mathbf{F}$:

$$R_\pm(T, \mathbf{F}) = k_{12}(T)\, \exp(\pm \mathbf{p}.\mathbf{F}/2\, k_B T)$$

$k_{12}(T) \equiv k_{12}(T, \mathbf{0})$ is the zero-field rate. $\mathbf{p}$ is an oscillator-associated dipole. The axial electric conductivity $\sigma_c(T)$ under the low-field condition ($\mathbf{p}\cdot\mathbf{F} \ll 2 k_B T$) for $\mathbf{F} \parallel$ c-axis:

$$\sigma_c(T) = Ne\, L\, k_{12}(T)\, (p / k_B T) \equiv Ne\, \mu(T)$$

The axial electric resistivity is:

$$\rho_c(T) = (k_B / Ne\, p\, L)\, [\, T / k_{12}{}^C(T)\,] \equiv P_c[T/k_{12}{}^C(T)]\ (\Omega\cdot cm).$$

Phonon-coupled rate. A theory for $k_{12}(T)$ is available. We used a harmonic-oscillator version:[6]

$$k_{12}(T) = 2\nu \sinh(h\omega/2 k_B T) \times \{\sum_{E_n \gg \varepsilon_B}(2[1-\exp(-2\pi\gamma_n)]/[2-\exp(-2\pi\gamma_n)])\exp(-E_n / k_B T) +$$

$$\sum_{E_n \ll \varepsilon_B}(2\pi\gamma_n{}^{2\gamma_n-1}\exp(-2\gamma_n)/[\Gamma(\gamma_n)]^2)(\pi[F_{nn}(q_0,q_C)/2^n n!]^2)\exp(-\varepsilon_R / h\omega)\exp(-E_n / k_B T)\}$$

$\nu$- renormalized vibrational frequency ($\omega = 2\pi\nu$), Q- configurational (mode) coordinate ($q = \sqrt{K/h\omega}Q$), $K = M\omega^2$ - stiffness, $F_{nn}$- quadratic form of Hermite polynomials:

$$F_{nn}(q_0,q_C) = 2q_0 H_n(q_C)H_n(q_C-2q_0) - 2nH_{n-1}(q_C)H_{n-1}(q_C-2q_0) + 2nH_n(q_C)H_{n-1}(q_C-2q_0).$$

$\gamma_n = (\varepsilon_{\alpha\beta}{}^2/8\, h\omega)[1/\sqrt{(\varepsilon_R\, |E_n-\varepsilon_C|)}]$ is Landau-Zener's parameter. $\varepsilon_B$, $\varepsilon_C$, and $\varepsilon_R$ are barrier height, crossover and reorganization energies, respectively. In our particular case, $q_C=0$, $q_0$ is as given below. Throughout, $h$ is Planck's constant $/ 2\pi$.

Electric resistivity. The electric resistivity $\rho_c(T)$ exhibits both observed features: it first drops steeply within the thermally-activated branch of $k_{12}(T)$ and then rises linearly with T as the field-coupling term becomes predominating. In Figure 1 our formula is compared with experimental $\rho_c(T)$ plots at various *x*. Our theory applies to $\rho_{ab}(T)$ data too. Fits in Figure 1 conform to the $\rho_{ab}(T)$ experimental points too, however, up to a larger misfit factor.

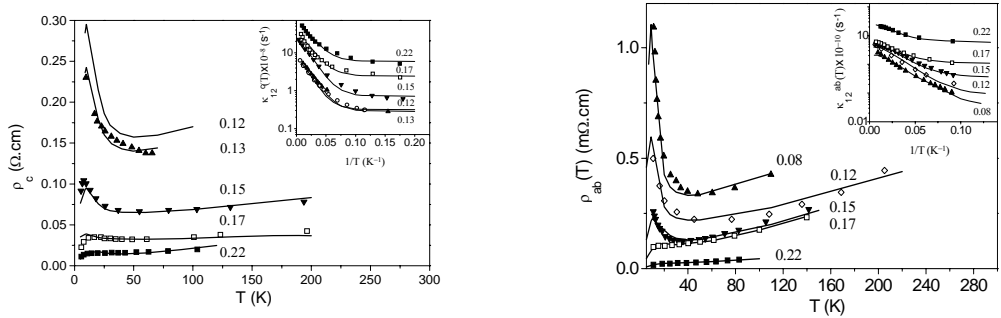

Figure 1. Temperature dependence of $\rho_c$ and $\rho_{ab}$: experimental data (symbols), theory (solid lines)

The agreement between experiment and theory suggests that fermionic excitations both in-plane and axial carry the normal state electric transport in a way controlled by scattering via small polarons formed as these carriers couple to vibrational modes. This conclusion is independent of the particular form of the local polaron potentials. We further extend the model so as to account for the site-splitting potentials at oxygen sites observed in experiments.

Site splitting potentials. Hamiltonian. The site-splitting analysis requires two narrow electronic bands and a mixing phonon field of the appropriate symmetry.[7] The phonon coupling splitting the site, the mixing mode is a symmetry-breaking vibration transforming according to an irreducible representation of the point group. If the latter group contains the spatial inversion, as at the O(A)-Cu(P)-O(A) molecule, the mixing mode will be odd parity for nearly-degenerate electronic bands composed of opposite-parity states (Pseudo-Jahn-Teller Effect). Such is the particular case of two oxygen-frame O(A) $2p_z$ orbitals (odd parity), on the one hand, and one in-plane Cu(P) orbital (even parity), on the other. The mixing vibration may be one of the $A_{2u}$ or the $E_u$ odd-parity modes, both symmetry breaking. For sufficiently strong coupling, octahedral (tetragonal) site symmetry will be lowered to that of an $A_{2u}$- or an $E_u$- deformed octahedron. An alternative case is that of two degenerate bands mixed by an even-parity symmetry-breaking phonon, e.g. of the $E_g$ symmetry (Dynamic-Jahn-Teller Effect). Mixing of Cu(P) $3d_z^2$ and $3d_{x^2-y^2}$ $e_g$- symmetry orbitals produces the axial lengthening (tetragonal $E_g$ distortion) of the $CuO_6$ octahedra.

The site-splitting Hamiltonian is:

$$H = \sum_{n,\mu} E_{n,\mu} a_{n,\mu}^+ a_{n,\mu} + \sum_{n,\mu} J_{n,\mu} a_{n,\mu}^+ (a_{n+1,\mu} + a_{n-1,\mu}) + \sum_{n\mu\nu} g^{n,\mu\nu} a_{n,\mu}^+ a_{n,\nu} (b_n^+ + b_n) + \sum_n (\hbar\omega_n) b_n^+ b_n$$

where the sums are: local energy, kinetic (hopping) energy, electron-phonon coupling energy, and phonon energy, respectively. n - site label, $\mu$, $\nu$- band labels ($\mu,\nu= 1,2$). $a_{n,\mu}^+$ ($a_{n,\mu}$) - fermion creation (annihilation) operators, $b_n^+(b_n)$ - boson creation (annihilation) operators, $E_{n,\mu}$ - local fermion energies, $J_{n,\mu}$ - fermion hopping energies, $g^{n,\mu\nu}$ - fermion-boson coupling constants, $\omega_n$ - bare phonon frequencies. Point-group symmetry requirement: $\Gamma_{ph} \subset \Gamma_1 \otimes \Gamma_2$.

Local Dynamics. To adiabatic approximation, the electron-phonon coupling is semiclassical as the mode coordinate $Q = \sqrt{\{\hbar\omega/K\}}(b_n^+ +b_n)$ is a c-number. The local Hamiltonian ($J_\mu = 0$) is:

$$H_{local} = \tfrac{1}{2}E_{12} (|2\rangle\langle2| - |1\rangle\langle1|) + G (|2\rangle\langle1| + |1\rangle\langle2|) + \tfrac{1}{2}( \mathbf{P}^2/M + KQ^2)$$

$K = M\omega^2$ - stiffness, M - reduced mass of vibrator, $G = g^{\mu\nu} \sqrt{\{K/(\hbar\omega)\}}$ - local coupling constant, $E_{12} = |E_2 - E_1|$ is the energy gap. In $\{|1\rangle,|2\rangle\}$ basis, the first-order-perturbation local energies are:

$$E_\pm = \tfrac{1}{2}(KQ^2 \pm \sqrt{\{(2GQ)^2 + E_{12}^2\}})$$

The upper adiabatic branch ($E_+$) is always minimal at Q = 0. The lower branch ($E_-$), minimal at Q = 0 for $E_{12} > 4 E_{JT}$, develops an instability (lateral minima at $\pm Q_0$) as the extremum at Q = 0 turns into a barrier maximum in between at $E_{12} < 4 E_{JT}$. $E_{JT} = G^2 / 2K$ is the Jahn-Teller energy,

$$Q_0 = \sqrt{\{(2 E_{JT} / K) (1 - \eta^2)\}},$$

($\eta = E_{12} / 4 E_{JT}$) is the distorted-configuration coordinate. The local polaron which forms at $\eta < 1$ is off-center relative to Q = 0, its configurational environment exhibiting the lower symmetry at $Q_0$. Will this symmetry feature withstand the opposite increased kinetic-energy trends of the finite $J_\mu$?

Hybridized O(A)-Cu(P)-O(A) molecule. A basic element of the $MO_6$ octahedra in cuprates is the O(A)-M(P)-O(A) molecule. Its energy spectrum is composed of: bonding (B), non-bonding (NB), and antibonding (AB) levels:[8]

$$\begin{aligned} E_B &= E_O + \tfrac{1}{2}\{\varepsilon_{O-M} - \sqrt{(\varepsilon_{O-M}^2 + 8 t_{M-O}^2)}\} \\ E_{NB} &= E_O \\ E_{AB} &= E_O + \tfrac{1}{2}\{\varepsilon_{O-M} + \sqrt{(\varepsilon_{O-M}^2 + 8 t_{M-O}^2)}\} \end{aligned}$$

Here $\varepsilon_{O-M} = |\varepsilon_M - \varepsilon_O|$ where $\varepsilon_O$ and $\varepsilon_M$ are the energies of the oxygen $|2p_z\rangle$ and metal $|3d_z^2\rangle$ orbitals, respectively, when these ions are far apart, $t_{M-O}$ is the hybridization (M-O hopping) energy. $E_O$ is an oxygen level doubly-degenerate in the absence of hybridization with the metal orbitals.

For $t_{M-O} \ll \varepsilon_{O-M}$, the hybridization brings about but a small splitting of the doublet levels. The eigenstates are:

$$\psi_O = 2^{-1/2} (|1\rangle + |2\rangle)$$
$$\psi_B = \sin\gamma |0\rangle + 2^{-1/2} \cos\gamma (|1\rangle - |2\rangle)$$
$$\psi_{AB} = \sin\beta |0\rangle + 2^{-1/2} \cos\beta (|1\rangle - |2\rangle)$$

where $|0\rangle$ is the $3d_z^2$ orbital state, $|1\rangle$ and $|2\rangle$ are $|2p_z\rangle$ orbitals of two lateral oxygens, while $\tan\beta = E_{AB} / \sqrt{2} t_{M-O}$, $\tan\gamma = E_B / \sqrt{2} t_{M-O}$.

Symmetry of O(A)-Cu(P)-O(A) molecule. The point group of the $CuO_6$ octahedron is $O_h$ in the absence of distortion.[7] The symmetry-breaking vibrations of O(A)-Cu(P)-O(A) are O(A) displacements mainly, even-parity $E_g$ and $T_{2g}$ and odd-parity $T_{1u}$ and $T_{2u}$. In a $CuO_6$ octahedron with tetragonal distortion, $T_{1u}$ splits to planar- $E_u$ and axial- $A_{2u}$.

The metal orbital $3d_z^2$ is $e_g$ symmetry relative to the inversion center Cu(P), the O(A)-Cu(P)-O(A) orbital frame vibrates as a $a_{2u}$ bi-orbital coupled to the $A_{2u}$ mode. A Pseudo-Jahn-Teller (PJT) mixing of $e_g$ and $a_{2u}$ by the $A_{2u}$ mode is conceivable at $t_{M-O} = 0$ leading to a configurational double-well instability of the ground-state doublet. The PJT mixing effect may be imprinted on the optical spectra. As hybridization is let go, the doublet splits and coupling to the $E_g$ axial mode is activated which results in a Dynamic-Jahn-Teller (DJT) effect relaxing the instability at the upper component while still leaving a configurational instability at the lower one.

For $e_g$ orbital symmetry, the symmetric quadrate $E_g \otimes E_g = E_g \oplus A_{2g}$ decomposes into symmetry-breaking and symmetry-retaining components. A due account of the point-group requires introducing two coordinates, $Q_E$ and $Q_A$, so that the interaction DJT Hamiltonian reads:

$$H_{int} = \sum_n g_E (a_{n\,2}^+ a_{n\,1} + a_{n\,1}^+ a_{n\,1}) Q_E + \sum_n g_A (a_{n\,2}^+ a_{n\,2} - a_{n\,1}^+ a_{n\,1}) Q_A$$

The conclusion that both PJT and DJT couplings comprise the electron-mode interactions at the $MO_6$ octahedra is essential for understanding the quasi-1D polaron propagation along c-axis.

Conclusion. It appears that the axial polaron transport in the cuprates is mainly controlled by scattering from the ground-state double-well instability. Our fitting parameters point to Jahn-Teller energies $\varepsilon_{JT} \sim 5$ meV at gap energies $\varepsilon_{gap} \sim 0.1$ meV. Apparently, hybridization of the eigenstates is essential for carrier transport: Taking into account the electron hopping, e.g. $t_{b1-a2} \sim 15$ meV,[9] the B-NB energy gap is $\varepsilon_{gap} = \frac{1}{2}|\Delta_{b1-a2} - \sqrt{\{\Delta_{b1-a2}^2 + 8t_{b1-a2}^2\}}| \sim 0.1$ meV for $\Delta_{b1-a2} \sim 2.4$ eV (cluster calculation for a O(A)-Cu(P)-O(A) molecule). This is the energy gap in the basis of hybridized symmetric and antisymmetric combinations of the $b_1$ and $a_2$ molecular orbitals of the $La_8CuO_6$ cluster. It now enters as an energy gap in the DJT mixing of hybridized levels by $A_{2g}$ and $E_g$ modes generating double wells and thereby a dynamic coupling which gives rise to the $\rho(T)$ temperature dependence.

Elsewhere, the semiconductor-like branch on the $\rho(T)$ experimental dependencies has been considered as more or less enigmatic. While a metallic temperature branch of the normal-state electric current has been anticipated, the thermally-activated branch is attributed to two factors: disorder-induced Anderson's localization, especially on chemical substitution ($\rho_{ab}$),[3] and to scattering by double-well potentials.[1]

References.